\documentclass[prl,twocolumn]{revtex4-1}
\usepackage{graphicx}
\usepackage{amsmath}
\usepackage{bm}
\usepackage{url}

\begin{document}

\title{Hierarchical freezing in a lattice model} 

\author{Travis W. Byington}
\affiliation{Physics Department, Duke University, Durham, NC 27708}

\author{Joshua E.~S.~Socolar}
\email[]{socolar@phy.duke.edu}
\affiliation{Physics Department, Duke University, Durham, NC 27708}


\date{\today}

\begin{abstract}
A certain two-dimensional lattice model with nearest and next-nearest neighbor interactions is known to have a limit-periodic ground state.
We show that during a slow quench from the high temperature, disordered phase, the ground state emerges through an infinite sequence of phase transitions.
We define appropriate order parameters and show that the transitions are related by renormalizations of the temperature scale.  As the temperature is decreased, sublattices with increasingly large lattice constants become ordered.  A rapid quench results in glass-like state due to kinetic barriers created by simultaneous freezing on sublattices with different lattice constants.
\end{abstract}


\maketitle 

A recent result in tiling theory~\cite{SocolarTaylorJCT} presents a new opportunity for studying the development of long-range order in a system with a non-periodic ground state.  A single hexagonal prototile has been shown to force a limit-periodic pattern (meaning a state made up of the union of an infinite number of periodic structures with lattice constants of ever-increasing sizes~\cite{GS,SocolarTaylorMI,Baake98}). 
The existence of a local Hamiltonian with a limit-periodic ground state composed of a single unit repeated in different orientations suggests that it may be possible for solid state materials, colloidal systems, or perhaps even collections of macroscopic units to realize structures of this type.  A similar type of complexity occurs in quasicrystals and has implications for electronic, photonic, elastic, and frictional properties.~\cite{janot}  Even in the absence of any physical examples of limit-periodic phases, however, the statistical mechanical properties of the new tiling model are remarkable and may serve as a conceptually useful interpolation between crystalline and glassy behavior. 

In this paper, we study the spontaneous formation of the limit-periodic structure.  Working with a substantially more complex tiling model with square symmetry, Mi\c{e}kisz showed that a series of partially ordered equilibrium states exist at nonzero temperatures in a system with a limit-periodic ground state~\cite{Miekisz1989}.  Here, we explicitly define order parameters for the transitions, present numerical evidence for ordering at slow cooling rates through an infinite sequence of phase transitions, and present a scaling theory of the transition hierarchy.  We also show that rapid quenching produces a state with frozen disorder resulting from the competition between two or more levels of the hierarchy.   

Our model Hamiltonian is based on rules for placing hexagonal tiles of the type shown in Fig.~\ref{fig:tiles}(a) on a close-packed lattice.   Each tile may be placed in any of the twelve orientations obtainable by rotations by $\pi/3$ and reflection.  For each nearest neighbor pair, we assign energy zero if the black stripe is continuous across their shared boundary and $\epsilon_1$ otherwise.  For each next-nearest neighbor pair, the energy is zero if the flags in their closest corners point in the same direction and $\epsilon_2$ otherwise.   Ref.~\cite{SocolarTaylorJCT} shows that for positive $\epsilon_1$ and $\epsilon_2$, the ground state of an infinite system is a zero-energy structure that contains black triangles with arbitrarily large side lengths.   We emphasize that the system is completely homogeneous, being a lattice of identical units.  

\section{Staggered tetrahedral order}
\label{sec:op}
The key to explaining the behavior of the lattice model is to focus on the patterns of (truncated) black triangles of different sizes in Fig.~\ref{fig:tiles}(c).  We refer to the smallest triangles as making up level 1, the next smallest as level 2, etc., and note that the edge of a triangle at level $n$ is a straight black stripe crossing $k_n-1$ tiles, where $k_n\equiv 2^{n-1}$.  At each level we have a periodic arrangement of triangles centered on the vertices of a honeycomb lattice.

\begin{figure*}[h]
\includegraphics[width=0.26\columnwidth]{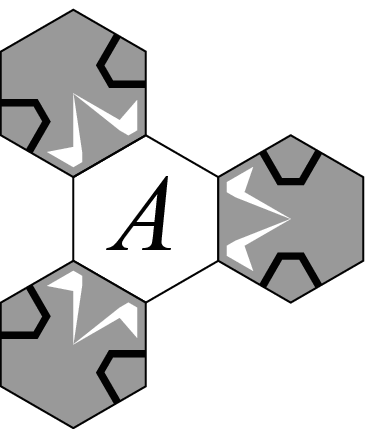}
\includegraphics[width=0.8\columnwidth]{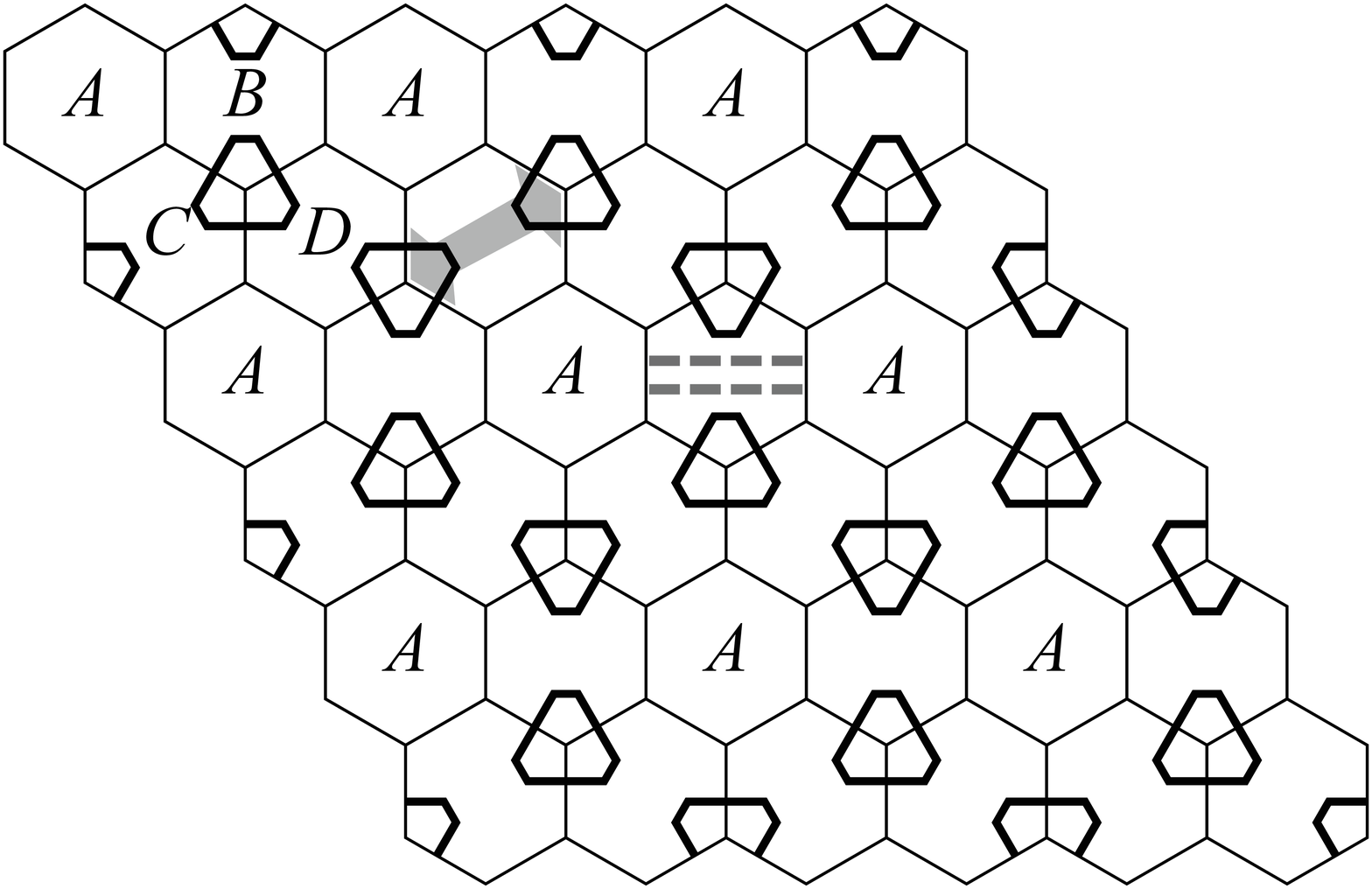}
\includegraphics[width=0.7\columnwidth]{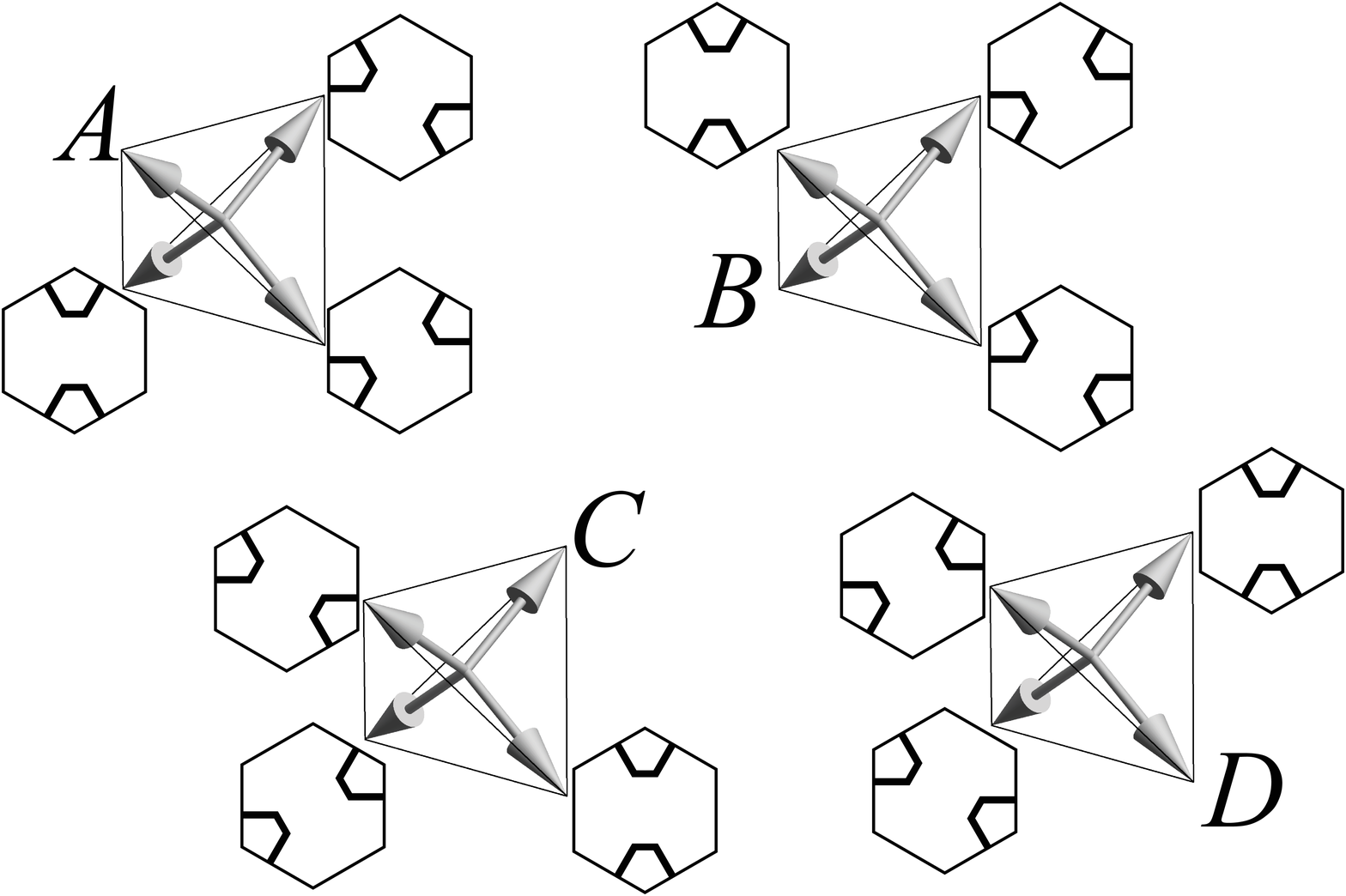}
\caption{The sublattices employed in the definition of the order parameter and the mapping from tile orientations to spin vectors.  The tiles of the $A$ sublattice are labeled along with one tile each of the $B$, $C$, and $D$ sublattices.  The figure shows the pattern of level-1 triangles formed when the non-contributing tiles are those of the $A$ sublattice.  For explanation of the cluster shown at left, the dashed lines, and grey bar, see text.}
\label{fig:op}
\end{figure*}

The level-1 pattern may form in four different ways.  One example is shown in Fig.~\ref{fig:op}, in which the tiles marked $A$ do not contribute at all to the triangles.  The other three are translations of this one, each with a different non-contributing sublattice ($B$, $C$, or $D$).  We associate with each tile $j$ a ``staggered tetrahedral spin'' vector ${\bm \sigma}_{1,j}=\vec{e}_X$, where $X$ indicates one of the four vertices of a reference tetrahedron.  (See Fig.~\ref{fig:op}.)  The spin of a tile is determined both by the orientation of the diameter joining its two black triangle corners and by the sublattice that it belongs to, according to the map shown at right in Fig.~\ref{fig:op}.   For example, a tile with corners aligned vertically and sitting on the $D$ sublattice is assigned $\sigma_{1}=\vec{e}_C$.  Note that specifying ${\bm \sigma}_{1,j}$ does not completely specify the orientation of tile $j$.  There are four consistent choices, corresponding to the two possible locations of the long black stripe and two possible orientations of the flag bar perpendicular to that stripe.   Note also that for any given tile, ${\bm \sigma}_{1}$ can take only three of the four possible values.

We define the average total spin ${\bm \sigma}_{1,{\rm tot}} \equiv \frac{1}{N}\sum_{j} {\bm \sigma}_{1,j}$, where $N$ is the number of tiles in the system.
The system exhibits tetrahedral symmetry in the following sense:  
for each configuration with a given ${\bm \sigma}_{1,{\rm tot}}$, there is another with identical energy having ${\bm \sigma}_{1,{\rm tot}}^{\prime}$ related to ${\bm \sigma}_{1,{\rm tot}}$ by an operation in the 24-element tetrahedral group $T_d$.  The mapping from operations on the lattice to elements of $T_d$ is given in Table~\ref{tab:group}.

\begin{table}[t]
\begin{center}
\begin{tabular}{|ccc|}
\hline
Lattice operation & \  & $T_d$ operation on ${\bm \phi}$ \\ [4pt]
\hline
\parbox[t]{1.7in}{\raggedright Rotation by $2\pi/3$ \\ $\quad$ about center of $X$} & $\rightarrow$ & \parbox[t]{1.5in}{\raggedright Rotation by $2\pi/3$ \\ $\quad$about ${\bm e}_X$} \\ [24pt] \hline
\parbox[t]{1.7in}{\raggedright Reflection through edge \\ $\quad$ shared by $X$ and $Y$} & $\rightarrow$ & \parbox[t]{1.5in}{\raggedright Reflection through \\ $\quad$ $({\bm e}_X,{\bm e}_Y)$ plane} \\  [24pt] \hline
\parbox[t]{1.7in}{\raggedright Translations taking \\ $\quad$ $X$ sublattice to $Y$ }& $\rightarrow$ & \parbox[t]{1.5in}{\raggedright Rotations by $\pi$ \\ $\quad$ about ${\bm e}_X+{\bm e}_Y$} \\ [24pt] \hline
\parbox[t]{1.7in}{\raggedright Rotation by $2\pi/3$ \\ $\quad$ followed by reflection} & $\rightarrow$ & \parbox[t]{1.5in}{\raggedright Rotary inversion} \\ [24pt]
\hline
\end{tabular}
\caption{Symmetry operations for the total staggered tetrahedral spin.  The left column specifies an operation on the 2D tiling pattern, where $X,Y \in  \left\{ A, B, C, D\right\}$ each represent a tile in the corresponding sublattice of Fig.~\ref{fig:op}.  The right column specifies 3D operations on the order parameter in terms of  the tetrahedral star of vectors ${\bm e}_X$, where $X$ is the label shown on Fig.~\ref{fig:op}.}
\label{tab:group}
\end{center}
\end{table}

Consider now the level-2 triangles.  If the level-1 structure is perfectly ordered, the corners of the level-2 triangles must come from the non-contributing sublattice at level 1.  That sublattice (e.g., all of the $A$'s in Fig.~\ref{fig:op}) is a precise copy of the original lattice (with distances scaled by a factor of 2), so we can define a new staggered tetrahedral spin ${\bm \sigma}_{2,{\rm tot}} \equiv \frac{1}{N/4}\sum_{j} {\bm \sigma}_{2,j}$, where the sum runs only over the sublattice of interest.  The construction can be repeated ad infinitum, with ${\bm \sigma}_{n,{\rm tot}}$ being well-defined if and only if the ordering on level $n-1$ is sufficiently strong to unambiguously specify which sublattice will order at level $n$.

\section{Frustration and order in quenches}
\begin{figure*}[h]
\begin{center}
\includegraphics[width={1.7\columnwidth}]{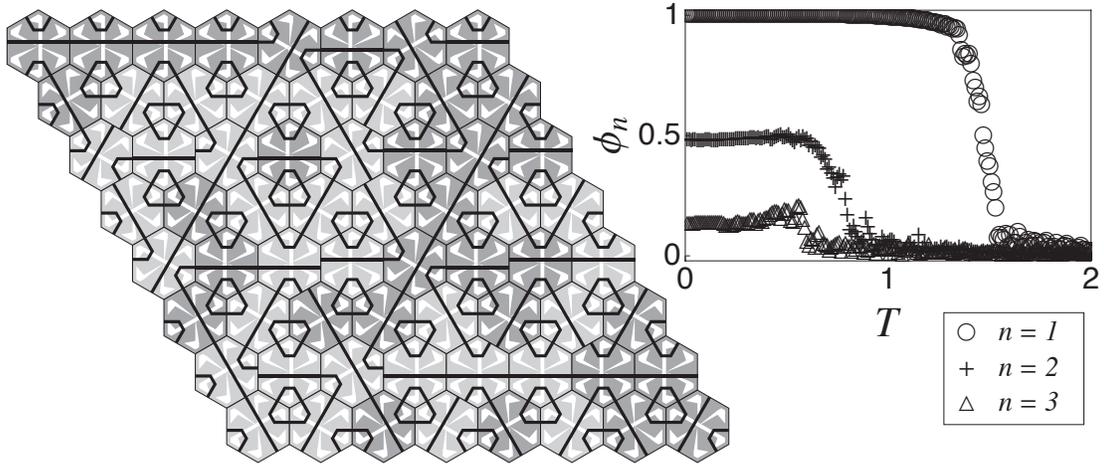}
\caption{Right: The behavior of the order parameters for a rapid quench.  Simulations were performed on a $64\times 64$ rhombic domain for $\epsilon_1 = \epsilon_2 = 1$ and $\tau = 120$.  Left:  The result of a rapid quench on an $8\times 8$ lattice showing the high density of defects that persists at long times.   There are no defects, however, in the level-1 structure, consistent with the high value of $\phi_1$.}
\label{fig:fastquench}
\end{center}
\end{figure*}

\begin{figure*}[h]
\begin{center}
\includegraphics[width={1.7\columnwidth}]{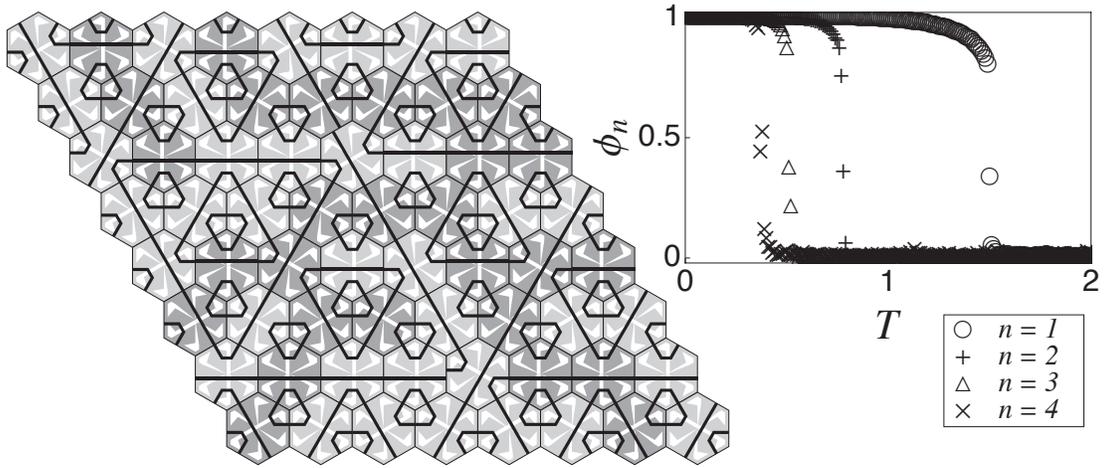}
\caption{Right: The behavior of the order parameters for a slow quench.  Simulations were performed on a $64\times 64$ rhombic domain for $\epsilon_1 = \epsilon_2 = 1$ and $\tau = 12\times 10^5$.  Left:  The result of a slow quench on an $8\times 8$ lattice showing the minimum possible number of defects.}
\label{fig:slowquench}
\end{center}
\end{figure*}

We define an order parameter for each level of structure:  $\phi_n \equiv \max [{\bm e}_X\cdot{\bm \sigma}_{n,{\rm tot}}]$, where $X\in\{A,B,C,D\}$ and ${\bm e}_X$ is a unit vector in the direction labeled ``$X$'' in Fig.~\ref{fig:op}.  We take the magnitude $\sigma$ of each spin to be $3/2$ so that the maximum $\phi_n$ is unity for all $n$.  
The maximum $\phi_1$ occurs if and only if the lattice of small triangles is perfectly ordered, regardless of the orientations of the tiles on the $X$ sublattice.  
Figs.~\ref{fig:fastquench} and~\ref{fig:slowquench}  show the behavior of $\phi_n$ with $n=1,2,3,4$ for a rapid quench and for a slow quench.  The simulations were done on rhombic domain with periodic boundary conditions.  Because the ground state is nonperiodic, there must be some mismatches at all times.  For domains of size $2^n\times 2^n$, the smallest possible number of mismatches is 4.
The simulations employed a standard Metropolis algorithm in which a random tile is chosen at each step and one of the twelve orientations of that tile is chosen randomly as a possible move.  $T$ is lowered by $\delta T = 0.01$ after every $N \tau$ steps, where $\tau$ controls the quench rate.

For the rapid quench (see Fig.~\ref{fig:fastquench}),  $\phi_1$ reaches its saturation value but $\phi_2$ does not.   The ordering on level 2 ceases to increase once the level-3 transition temperature is reached, at which point interactions associated with level 3 prevent further ordering of level 2.
For the slow quench, on the other hand, Fig.~\ref{fig:slowquench} shows a sequence of clear transitions.  The number of transitions that can be observed is limited by the system size.  At level 5 (not shown), the edge length of a black triangle is 15 tiles, so our $64\times 64$ system is only $4\times 4$ at level 5.  

Because the symmetry of the order parameter is the same as that of a 4-state Potts model, we expect the transition to fall in the same universality class, which has an order parameter exponent  $\beta = 1/12$.~\cite{WuRMP}  Preliminary numerical investigations are consistent with a very sharp second order transition, having not revealed any hysteresis in a cooling and heating cycle.  None of the analysis below depends on the critical behavior or even on the order of the transition.

\section{Scaling theory}
\label{sec:theory}
The system is considered fully ordered at level $n$ when all of the level-$n$ triangle corners occur in the pattern of Fig.~\ref{fig:tiles}(c).  In the ordered pattern, at each level, flags in the tiles surrounding a non-contributing tile also form triangular structures of the type shown at left in Fig.~\ref{fig:op}.  The long black stripes and the flag bars perpendicular to them, however, need not be in their ground state orientations.  Their role is only to mediate the interactions between the triangle corners.  We will see below that if the triangle corners at all levels $m < n$ are assumed to be perfectly ordered and immovable, the level-$n$ dynamics are identical to the level-1 dynamics at a rescaled temperature (and possibly a rescaled $\epsilon_2$). 
We discuss first the special case $\epsilon_1 = \epsilon_2 \equiv \epsilon$.

The partition function for the level-$n$ system can be expressed as follows.  A configuration of the system can be specified by giving, for every edge of each hexagon, the location of the black stripe meeting that edge and, for every vertex, the orientation of the flag at that vertex.  A configuration is allowed if and only if the specification for every hexagon corresponds to some orientation of the tile decoration of Fig.~\ref{fig:tiles}(a), independent of whether neighboring tiles match properly.  At level $n$, the contribution to the partition function from a given configuration of the triangle corners is a product of contributions from all of the nearest and next-nearest neighbor bonds between the corner tiles.  Each of these bonds consists of $k_n-1$ tiles, with black stripes joining nearest neighbors of the level-$n$ sublattice and flag bars joining next-nearest neighbors.  Fig.~\ref{fig:op} shows an example of each type of bond for the $n=2$ case.  The dashed grey lines show the two possible locations of the stripe on a triangle edge joining the corners that would be present on the $A$ sublattice.  The grey diagonal shows the two possibe orientations of the flag bar that forms a triangle edge for next-nearest neighbor interactions between two $A$ tiles.  

\begin{figure*}[h]
\begin{center}
\includegraphics[width={1.7\columnwidth}]{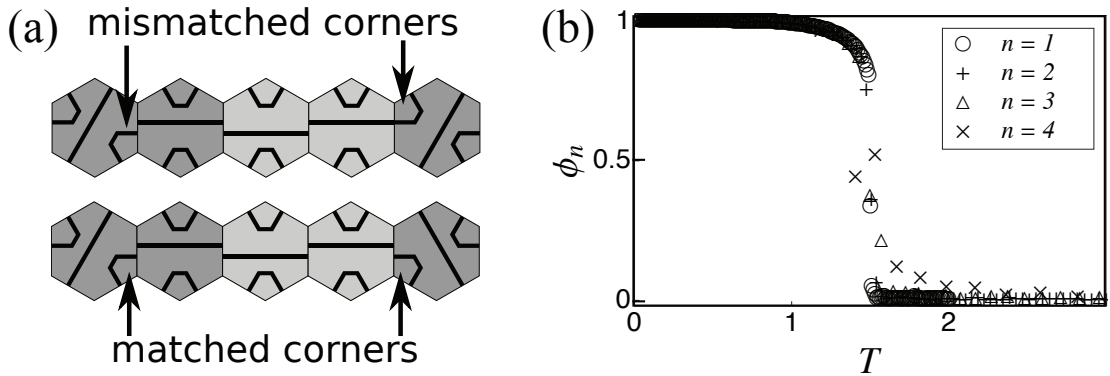}
\caption{(a) Matched and mismatched corner configurations, shown here for level-3 triangle edges.  (b) Scaling collapse of Fig.~\ref{fig:slowquench} data. The deviation of the level-4 points on the right is due to the finite size of the system.}
\label{fig:scaling}
\end{center}
\end{figure*}

The full partition function for the level-$n$ system is
\begin{equation}
Z_{n}(T) = \sum_{\rm configurations} \left(\prod_{\rm bonds}\zeta_{n,b}\right)\,,
\end{equation}
where $b\in \{{\rm odd, even}\}$ represents the state of a given bond (mismatched or matched) in the given configuration.
Each edge is effectively a 1D Ising system with $k_n$ possible mismatches (see Fig.~\ref{fig:scaling}(a)), which gives
\begin{eqnarray}
\zeta_{n, {\rm odd}} & = & \frac{1}{2}\left[\left(1+e^{-\epsilon/T}\right)^{k_n} - \left(1-e^{-\epsilon/T}\right)^{k_n}\right]\,; \nonumber \\
\zeta_{n, {\rm even}} & = &\frac{1}{2}\left[\left(1+e^{-\epsilon/T}\right)^{k_n} + \left(1-e^{-\epsilon/T}\right)^{k_n}\right] \,.
\end{eqnarray}  


Now because the configuration sums are identical for all levels, the behavior of the system at level $n$ at some temperature $T_n$ will be identical to that for level 1 at temperature $T_1$ if
$\zeta_{n,{\rm odd}}(T_n)  =  \alpha\, \zeta_{1,{\rm odd}}(T_1)$ 
and
$\zeta_{n,{\rm even}}(T_n)  =  \alpha\, \zeta_{1,{\rm even}}(T_1)$
for some normalization constant $\alpha$. 
Eliminating $\alpha$ from these equations yields, after some algebra,
\begin{equation}
\label{eqn:scaling}
\tanh\left(\frac{\epsilon}{2T_1}\right)= \left[\tanh\left(\frac{\epsilon}{2T_n}\right)\right]^{k_n}
\end{equation}
or, equivalently, for all $n$,
\begin{equation}
\label{eqn:scaling2}
\tanh\left(\frac{\epsilon}{2T_n}\right)= \left[\tanh\left(\frac{\epsilon}{2T_{n+1}}\right)\right]^2\,.
\end{equation}
From Eq.~\eqref{eqn:scaling}, the transition temperature for large $n$ is  
\begin{equation}
T_{c;n} = \frac{1}{n\log 2 - \log\left[\log\left[\coth(\frac{\epsilon}{2 T_{c;1}})\right]\right]} + {\cal O}\left(\frac{n}{2^{2n}}\right)\,. 
\label{eqn:Tcn}
\end{equation}

One might worry that Eq.~\eqref{eqn:Tcn} will break down because $\phi_n$ may not be fully saturated at $T_{c;n+1}$.  Nevertheless,
as shown in Fig.~\ref{fig:scaling}, we obtain an excellent data collapse for several levels by plotting $\phi_n(T_n)$ as a function of $T_1(T_n)$.
(Note:  The $\times$'s on the high temperature side of the transition represent finite-size fluctuations and correspond to projections onto different tetrahedral vectors.)
The collapse shows that the residual disorder after the transition at level $n$ has a very small effect on the transition at level $n+1$.  For $n=1$, the data for Fig.~\ref{fig:slowquench} show clearly that $\phi_1$ is very close to unity ($\approx 0.999$) at the critical temperature for $\phi_2$.  The scaling argument takes this value to be exactly unity, in which case Eq.~\eqref{eqn:scaling2} leads to $\phi_n(T_{c;n+1}) = \phi_1(T_{c;2})$ for all $n$.  If the small deviation from unity causes the scaling to break down at large $n$, it appears that the difficulty will only emerge at lattice sizes too large to be probed computationally.

The analysis is more complicated for the generic case $\epsilon_2 < \epsilon_1$, but the essential phenomenology appears to be the same.
Let $T_{c;n}(\epsilon_1,\epsilon_2)$ be the level-$n$ transition temperature.  Assuming that deviations discussed in the previous paragraph can still be neglected, $T_{c;n}$ must be an increasing function of each of its arguments.  Thus we must have $\epsilon_2/\epsilon_1<T_{c;n}(\epsilon_1,\epsilon_2)/T_{c;n}(\epsilon_1,\epsilon_1)<1$, where the first inequality follows from the fact that a simple rescaling of all energies gives $T_{c;n}(\epsilon_2,\epsilon_2) = (\epsilon_2/\epsilon_1) T_{c;n}(\epsilon_1,\epsilon_1)$.

\section{Discussion}
\label{sec:discussion}
\begin{figure*}[h]
\begin{center}
\includegraphics[width=1.7\columnwidth]{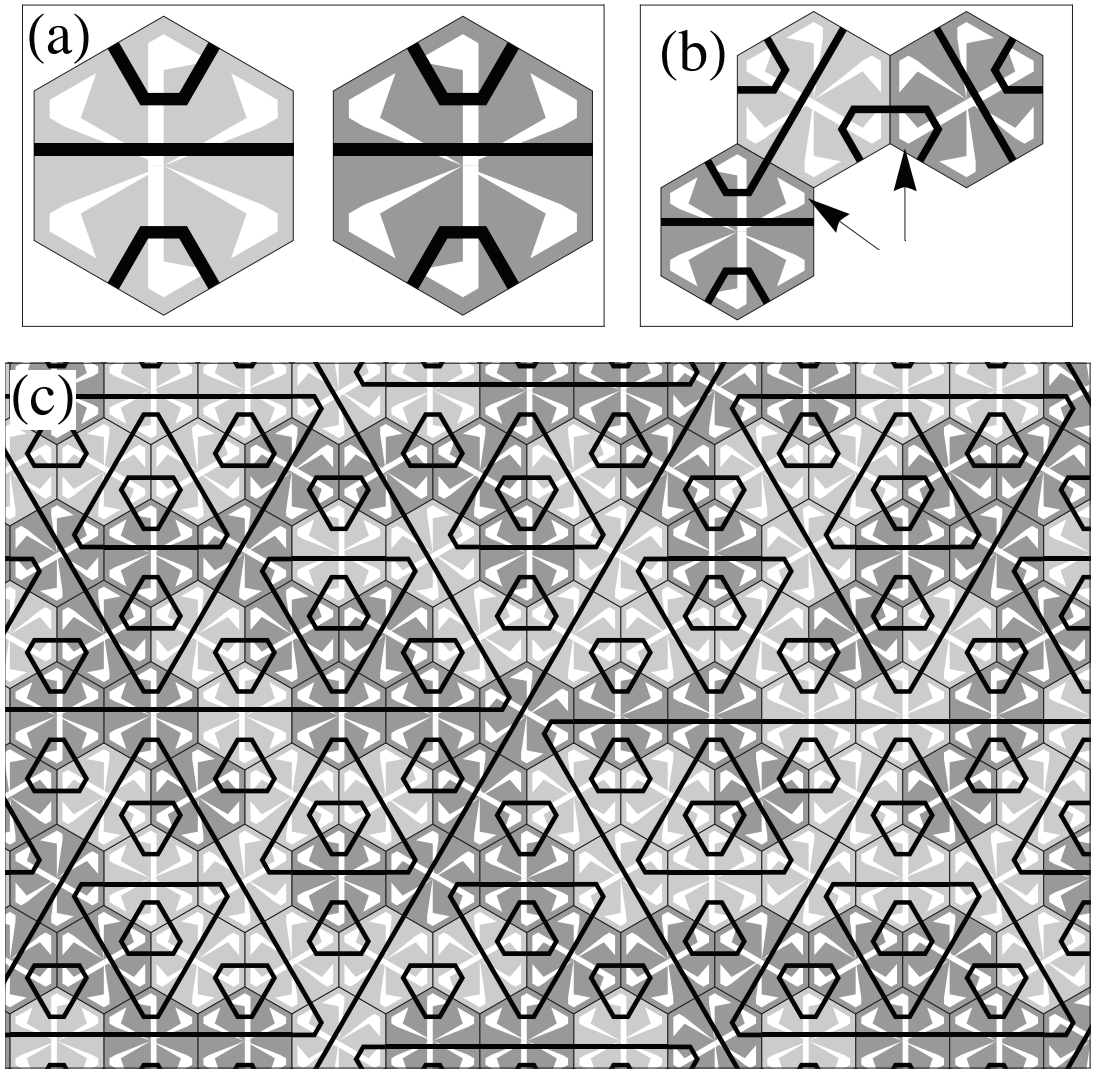}
\caption{The hexagonal prototile and its mirror image.  (a) The two tiles  are related by reflection about a vertical line.  (b)  For zero energy, adjacent tiles must form continuous black stripes and flag decorations at opposite ends of a tile edge (as indicated by the arrows) must point in the same direction.   (c) A portion of an infinite tiling that has zero energy.}
\label{fig:tiles}
\end{center}
\end{figure*}

The ground state of the tiling model is discussed in detail in Ref.~\cite{SocolarTaylorJCT}.  The complexity of the structure is best revealed by the pattern of tile parities (the grey and white in Fig.~\ref{fig:tiles}(c)).  
We have shown here that the model also has remarkable properties at finite temperatures.  First, it exhibits a hierarchy of distinct thermodynamic phases, with each successive one corresponding to formation of an ordered lattice with a lattice constant twice as large as the last.  Second, each transition is unusual in that the low temperature phase leaves one quarter of the tiles as ``rattlers'' with undetermined orientations.  Finally, the kinetics of ordering become frustrated if the quench rate is fast enough that the temperature drops below $T_{c;n}$ before $\phi_{n-1}$ has reached a sufficiently high value.  The latter effect is reminiscent of glass formation, where lower energy states can be reached through slower quenching but any nonzero quench rate eventually leads to trapping in a nonequilibrium configuration.~\cite{stillinger01}  The present model shows that this can happen in the context of a series of transitions that each establish true long-range order.

Many questions remain concerning the precise nature of the staggered tetrahedral phase transition, the possibility of reaching the ground state via growth from a small seed, the geometry and energetics of topological defects in this system, and the scaling of the time required for ordering at each level.  

To date, we know of no examples of spontaneous emergence of limit-periodic order in a real system.  The present work suggests that such systems may exist and exhibit novel properties.  The hexagonal tiling consists of a single unit that may be realized as a cluster of atoms, a surface pattern on a colloidal particle, or even a micro-machined brick.  It is also possible that such systems have already been observed but not recognized because they present as ``poor" crystals in diffraction studies. (See Ref.~\cite{Baake11} for a discussion of diffraction from a limit-periodic structure.)  Systems of interest may be two-dimensional monolayers on a flat substrate, surface reconstructions on crystals, or bulk phases.  See Refs.~\cite{SocolarTaylorJCT} for a 3D tile that enforces the present structure through its shape alone by allowing direct contact between next-nearest neighbors. 

Mi\c{e}kisz's study~\cite{Miekisz1990} based on an extension of Robinson's set of six Wang tiles~\cite{Robinson} indicates that a limit-periodic ground state with square symmetry should show a similar sequence of transitions.  See~\cite{GoodmanStrauss99a} and~\cite{GoodmanStrauss99b} for 2D and 3D examples of limit-periodic tilings with square symmetry composed of only two types of tiles, which may be more amenable to the type of analysis performed above.

\begin{acknowledgments}
We thank Patrick Charbonneau, Tom Lubensky, David Huse, and Giulio Biroli for helpful conversations.
This work was supported by an Undergraduate Summer Fellowship from the Duke Physics Department and the NSF Research Triangle MRSEC (DMR-1121107).
\end{acknowledgments}

\bibliography{tiling}

\begin{thebibliography}{13}%
\makeatletter
\providecommand \@ifxundefined [1]{%
 \@ifx{#1\undefined}
}%
\providecommand \@ifnum [1]{%
 \ifnum #1\expandafter \@firstoftwo
 \else \expandafter \@secondoftwo
 \fi
}%
\providecommand \@ifx [1]{%
 \ifx #1\expandafter \@firstoftwo
 \else \expandafter \@secondoftwo
 \fi
}%
\providecommand \natexlab [1]{#1}%
\providecommand \enquote  [1]{``#1''}%
\providecommand \bibnamefont  [1]{#1}%
\providecommand \bibfnamefont [1]{#1}%
\providecommand \citenamefont [1]{#1}%
\providecommand \href@noop [0]{\@secondoftwo}%
\providecommand \href [0]{\begingroup \@sanitize@url \@href}%
\providecommand \@href[1]{\@@startlink{#1}\@@href}%
\providecommand \@@href[1]{\endgroup#1\@@endlink}%
\providecommand \@sanitize@url [0]{\catcode `\\12\catcode `\$12\catcode
  `\&12\catcode `\#12\catcode `\^12\catcode `\_12\catcode `\%12\relax}%
\providecommand \@@startlink[1]{}%
\providecommand \@@endlink[0]{}%
\providecommand \url  [0]{\begingroup\@sanitize@url \@url }%
\providecommand \@url [1]{\endgroup\@href {#1}{\urlprefix }}%
\providecommand \urlprefix  [0]{URL }%
\providecommand \Eprint [0]{\href }%
\providecommand \doibase [0]{http://dx.doi.org/}%
\providecommand \selectlanguage [0]{\@gobble}%
\providecommand \bibinfo  [0]{\@secondoftwo}%
\providecommand \bibfield  [0]{\@secondoftwo}%
\providecommand \translation [1]{[#1]}%
\providecommand \BibitemOpen [0]{}%
\providecommand \bibitemStop [0]{}%
\providecommand \bibitemNoStop [0]{.\EOS\space}%
\providecommand \EOS [0]{\spacefactor3000\relax}%
\providecommand \BibitemShut  [1]{\csname bibitem#1\endcsname}%
\let\auto@bib@innerbib\@empty
\bibitem [{\citenamefont {Socolar}\ and\ \citenamefont
  {Taylor}(2011)}]{SocolarTaylorJCT}%
  \BibitemOpen
  \bibfield  {author} {\bibinfo {author} {\bibfnamefont {J.~E.~S.}\
  \bibnamefont {Socolar}}\ and\ \bibinfo {author} {\bibfnamefont {J.~M.}\
  \bibnamefont {Taylor}},\ }\href@noop {} {\bibfield  {journal} {\bibinfo
  {journal} {Journal of Combinatorial Theory: Series A}\ }\textbf {\bibinfo
  {volume} {118}},\ \bibinfo {pages} {2207} (\bibinfo {year}
  {2011})}\BibitemShut {NoStop}%
\bibitem [{\citenamefont {Gr\"{u}nbaum}\ and\ \citenamefont
  {Shephard}(1986)}]{GS}%
  \BibitemOpen
  \bibfield  {author} {\bibinfo {author} {\bibfnamefont {B.}~\bibnamefont
  {Gr\"{u}nbaum}}\ and\ \bibinfo {author} {\bibfnamefont {G.~C.}\ \bibnamefont
  {Shephard}},\ }\href@noop {} {\emph {\bibinfo {title} {Tilings and
  patterns}}}\ (\bibinfo  {publisher} {W. H. Freeman \& Co.},\ \bibinfo
  {address} {New York, NY, USA},\ \bibinfo {year} {1986})\BibitemShut {NoStop}%
\bibitem [{\citenamefont {Socolar}\ and\ \citenamefont
  {Taylor}(2012)}]{SocolarTaylorMI}%
  \BibitemOpen
  \bibfield  {author} {\bibinfo {author} {\bibfnamefont {J.~E.~S.}\
  \bibnamefont {Socolar}}\ and\ \bibinfo {author} {\bibfnamefont {J.~M.}\
  \bibnamefont {Taylor}},\ }\href@noop {} {\bibfield  {journal} {\bibinfo
  {journal} {The Mathematical Intelligencer}\ }\textbf {\bibinfo {volume} {34}}
  (\bibinfo {year} {2012})},\ \bibinfo {note} {online at
  http://www.springerlink.com/content/v06145n476l13xp0/}\BibitemShut {NoStop}%
\bibitem [{\citenamefont {Baake}\ \emph {et~al.}(1998)\citenamefont {Baake},
  \citenamefont {Moody},\ and\ \citenamefont {Schlottmann}}]{Baake98}%
  \BibitemOpen
  \bibfield  {author} {\bibinfo {author} {\bibfnamefont {M.}~\bibnamefont
  {Baake}}, \bibinfo {author} {\bibfnamefont {R.}~\bibnamefont {Moody}}, \ and\
  \bibinfo {author} {\bibfnamefont {M.}~\bibnamefont {Schlottmann}},\
  }\href@noop {} {\bibfield  {journal} {\bibinfo  {journal} {Journal of Physics
  A: Mathematical and General}\ }\textbf {\bibinfo {volume} {31}},\ \bibinfo
  {pages} {5755} (\bibinfo {year} {1998})},\ \bibinfo {note} {{S}ee also
  D.~Frettl{\"o}h, PhD Thesis, Universit{\"a}t Dortmund (2002).}\BibitemShut
  {Stop}%
\bibitem [{\citenamefont {Janot}(1997)}]{janot}%
  \BibitemOpen
  \bibfield  {author} {\bibinfo {author} {\bibfnamefont {C.}~\bibnamefont
  {Janot}},\ }\href@noop {} {\emph {\bibinfo {title} {Quasicrystals: a
  primer}}}\ (\bibinfo  {publisher} {Oxford University Press.},\ \bibinfo
  {address} {New York},\ \bibinfo {year} {1997})\BibitemShut {NoStop}%
\bibitem [{\citenamefont {Mi\c{e}kisz}(1989)}]{Miekisz1989}%
  \BibitemOpen
  \bibfield  {author} {\bibinfo {author} {\bibfnamefont {J.}~\bibnamefont
  {Mi\c{e}kisz}},\ }\href@noop {} {\bibfield  {journal} {\bibinfo  {journal}
  {Phys.~Lett.~A}\ }\textbf {\bibinfo {volume} {138}},\ \bibinfo {pages} {415}
  (\bibinfo {year} {1989})}\BibitemShut {NoStop}%
\bibitem [{\citenamefont {Wu}(1982)}]{WuRMP}%
  \BibitemOpen
  \bibfield  {author} {\bibinfo {author} {\bibfnamefont {F.~Y.}\ \bibnamefont
  {Wu}},\ }\href@noop {} {\bibfield  {journal} {\bibinfo  {journal} {Reviews of
  Modern Physics}\ }\textbf {\bibinfo {volume} {54}},\ \bibinfo {pages} {235}
  (\bibinfo {year} {1982})}\BibitemShut {NoStop}%
\bibitem [{\citenamefont {Debenedetti}\ and\ \citenamefont
  {Stillinger}(2001)}]{stillinger01}%
  \BibitemOpen
  \bibfield  {author} {\bibinfo {author} {\bibfnamefont {P.~G.}\ \bibnamefont
  {Debenedetti}}\ and\ \bibinfo {author} {\bibfnamefont {F.~H.}\ \bibnamefont
  {Stillinger}},\ }\href {http://dx.doi.org/10.1038/35065704} {\bibfield
  {journal} {\bibinfo  {journal} {Nature}\ }\textbf {\bibinfo {volume} {410}},\
  \bibinfo {pages} {259} (\bibinfo {year} {2001})}\BibitemShut {NoStop}%
\bibitem [{\citenamefont {Baake}\ and\ \citenamefont {Grimm}(2011)}]{Baake11}%
  \BibitemOpen
  \bibfield  {author} {\bibinfo {author} {\bibfnamefont {M.}~\bibnamefont
  {Baake}}\ and\ \bibinfo {author} {\bibfnamefont {U.}~\bibnamefont {Grimm}},\
  }\href {\doibase 10.1080/14786435.2010.508447} {\bibfield  {journal}
  {\bibinfo  {journal} {Philosophical Magazine}\ }\textbf {\bibinfo {volume}
  {91}},\ \bibinfo {pages} {2661} (\bibinfo {year} {2011})}\BibitemShut
  {NoStop}%
\bibitem [{\citenamefont {Mi\c{e}kisz}(1990)}]{Miekisz1990}%
  \BibitemOpen
  \bibfield  {author} {\bibinfo {author} {\bibfnamefont {J.}~\bibnamefont
  {Mi\c{e}kisz}},\ }\href@noop {} {\bibfield  {journal} {\bibinfo  {journal}
  {J.~Stat.~Phys.}\ }\textbf {\bibinfo {volume} {58}},\ \bibinfo {pages} {1137}
  (\bibinfo {year} {1990})}\BibitemShut {NoStop}%
\bibitem [{\citenamefont {Robinson}(1971)}]{Robinson}%
  \BibitemOpen
  \bibfield  {author} {\bibinfo {author} {\bibfnamefont {R.}~\bibnamefont
  {Robinson}},\ }\href@noop {} {\bibfield  {journal} {\bibinfo  {journal}
  {Inventiones Mathematicae}\ }\textbf {\bibinfo {volume} {12}},\ \bibinfo
  {pages} {177} (\bibinfo {year} {1971})}\BibitemShut {NoStop}%
\bibitem [{\citenamefont
  {Goodman-Strauss}(1999{\natexlab{a}})}]{GoodmanStrauss99a}%
  \BibitemOpen
  \bibfield  {author} {\bibinfo {author} {\bibfnamefont {C.}~\bibnamefont
  {Goodman-Strauss}},\ }\href {\doibase DOI: 10.1006/eujc.1998.0281} {\bibfield
   {journal} {\bibinfo  {journal} {European Journal of Combinatorics}\ }\textbf
  {\bibinfo {volume} {20}},\ \bibinfo {pages} {375 } (\bibinfo {year}
  {1999}{\natexlab{a}})}\BibitemShut {NoStop}%
\bibitem [{\citenamefont
  {Goodman-Strauss}(1999{\natexlab{b}})}]{GoodmanStrauss99b}%
  \BibitemOpen
  \bibfield  {author} {\bibinfo {author} {\bibfnamefont {C.}~\bibnamefont
  {Goodman-Strauss}},\ }\href {\doibase DOI: 10.1006/eujc.1998.0282} {\bibfield
   {journal} {\bibinfo  {journal} {European Journal of Combinatorics}\ }\textbf
  {\bibinfo {volume} {20}},\ \bibinfo {pages} {385 } (\bibinfo {year}
  {1999}{\natexlab{b}})}\BibitemShut {NoStop}%
\end{thebibliography}%


\end{document}